# Recovering the History of Informed Consent for Data Science and Internet Industry Research Ethics

Elaine Sedenberg[1]
School of Information
University of California, Berkeley

Anna Lauren Hoffmann
School of Information
University of California, Berkeley

**Abstract**

Respect for persons is a cornerstone value for any conception of research ethics—though how to best realize respect in practice is an ongoing question. In the late 19th and early 20th centuries, "informed consent" emerged as a particular way to operationalize respect in medical and behavioral research contexts. Today, informed consent has been challenged by increasingly advanced networked information and communication technologies (ICTs) and the massive amounts of data they produce—challenges that have led many researchers and private companies to abandon informed consent as untenable or infeasible online.

Against any easy dismissal, we aim to recover insights from the history of informed consent as it developed from the late 19th century to today. With a particular focus on the United States policy context, we show how informed consent is not a fixed or monolithic concept that should be abandoned in view of new data-intensive and technological practices, but rather it is a mechanism that has always been fluid—it has constantly evolved alongside the specific contexts and practices it is intended to regulate. Building on this insight, we articulate some specific challenges and lessons from the history of informed consent that stand to benefit current discussions of informed consent and research ethics in the context of data science and Internet industry research.

**Introduction**

Respect for persons is a cornerstone value for any conception of research ethics—though how to best realize respect in practice is an ongoing question. In the late 19th and early 20th centuries, "informed consent" emerged as a particular way to operationalize respect in medical, behavioral, and social scientific research contexts. Today, these two concepts—respect for persons and informed consent—have become increasingly intertwined so as to be virtually indistinguishable (Corrigan, 2003; Dickert, 2009). At the same time, informed consent has been challenged—if not outright undermined—by research activities enabled by increasingly advanced networked information and communication technologies (ICTs) and the massive quantities of social and individually identifiable data they can produce. The rise of "big data" has raised new and compounded existing questions around the ethical treatment of human subjects in research, as legacy models forged in medical and behavioral research contexts are not easily applied to data scientific practice (Metcalf & Crawford, 2016). Consequently, many researchers and private companies have ignored or outright abandoned core research ethics ideas like "informed consent" as untenable, relying instead on notice and consent practices which do not share the same grounding in research ethics or even moral theory generally. This results in often opaque or impenetrable privacy policies and data use notices or—worse—no notice at all. (Jensen & Potts, 2004; Pollach, 2007; Hayden, 2012).

---

[1] *Author names are listed in reverse alphabetical order; both authors contributed equally to this article.*

Despite the seeming newness of much "big data" research, however, the future of ethical research requires, as Metcalf and Crawford (2016) argue, that data ethics be situated "within a historical and discursive analysis of the core concepts and norms of research ethics in general" (p. 2). To this end, we trace the development of "informed consent" in United States legal and policy regulations in order to better articulate its value for thinking about ethical issues with regard to data, technology, and research today. Importantly, we show how informed consent is not a fixed or monolithic concept, but a mechanism that has always been fluid, constantly evolving alongside the specific political contexts and research practices it has intended to regulate.

We begin with a comprehensive discussion of the development of informed consent as a legal, policy, and practical mechanism for operationalizing certain core values (like respect) and protecting human research subjects accordingly. From there, we detail the landscape of past and future challenges posed by forms of research conducted outside of medical and largely academic contexts as well as the new challenges posed by the rise of industry research and data-intensive ICTs. Finally, we draw on both the history and recent challenges to argue that a misguided focus on "informed consent" as a static or monolithic value distracts from a more robust debate about operationalizing the value of respect in view of specific, contextual features of data science. In so doing, we show how the history of informed consent can help shed light on developments and arguments relevant for thinking about data and research ethics today.

**The History of Informed Consent in the United States and Beyond**

***Consent and Voluntariness in Early Human Subjects Research Ethics***

Systematic scientific experimentation on human subjects was rare and isolated prior to the late 19th century. Early medical efforts were mostly therapeutic and intended to directly benefit an individual—only in rare cases was it intended to add to medical knowledge broadly. Accordingly, moral and ethical frameworks governing the actions of doctors (like the Hippocratic Oath) did not address experimentation; instead, they emphasized using all knowledge and remedies to the best extent possible in order to provide the best possible care for patients (Faden & Beauchamp, 1986). Despite a lack of attention to experimentation, however, early medical and bioethics discussions do contain precursors to the the idea of informed consent. In 1891, for example, a Prussian minister released a directive that tuberculin should not be administered as treatment for tuberculosis "against the patient's will" (Vollmann & Winau, 1996, p. 1445). This attention to patient consent during a therapeutic intervention for what would otherwise still be a fatal illness was an important step toward respecting patient autonomy (Vollmann & Winau, 1996).

Generalizable, non-therapeutic medical research emerged in the late 19th and early 20th century as doctors began testing preventive treatments on individuals. The expansion of medical sciences at this time sparked a bevy of chemical and therapeutic treatment options and uncovered potentially novel remedies with unknown risks and benefits, prompting doctors and early medical researchers to systematically compare treatments on human subjects. In 1898, for example, the German physician Albert Neisser began injecting sex workers with serum from syphilis patients as an experimental method for prevention—albeit without their consent (Vollmann & Winau, 1996). In 1900, partly as a response to the Neisser case, a Prussian directive issued to hospitals and clinics stated that medical interventions for any purpose other than diagnosis, healing, and immunisation must obtain "unambiguous consent" from patients after "proper explanation of the possible negative consequences" of the intervention (Vollmann & Winau, 1996).

In Germany in 1932, the Reich minister of the interior—responding to negative press and within the context of wider political reform in the country—issued detailed guidelines for both new therapies and nontherapeutic research (i.e., human experimentation) (Ghooi, 2011). These regulations were based on principles of beneficence, nonmaleficence, and patient autonomy. The requirements stated that a "new therapy may be applied only if consent or proxy consent has been given in a clear and undebatable manner following appropriate information" (1931 German Guidelines, 1980). In particular, the guidelines required written consent for all human experimentation. This early German policy established the idea of consent as a legal requirement and offered some guidelines for promoting values of beneficence, nonmaleficence, and autonomy in research. It is also notable that these early guidelines hint at something like "informed" consent—that is, consent as something that must follow "appropriate information."

These progressive and proactive research ethics policies, however, would soon fall by the wayside, ignored in the face of unbounded Nazi medical research and experimentation. Though there were many other cases of ethically questionable medical experimentation on human subjects in the early 20th century and up through World War II—namely in malarial treatments (Masterson, 2014, p. 162), the beginning of the Tuskegee syphilis study (Tuskegee Timeline, 2016), wartime starvation (Baker & Keramidas, 2013), and chemical warfare studies (Smith, 2008)—a significant amount of research ethics attention is paid to Nazi experimentation and the post-war Nuremberg Trials. Widespread media and global attention on the unsettling potential and costs of the kind of unadulterated experimentation undertaken by the Nazi regime spurred the worldwide policy community to establish legal and ethical guidelines to keep pace with scientific advancement.

The result of these efforts was the Nuremberg Code, drafted at the conclusion of the Nazi doctors' trial in 1947 (Hayden, 2012).[2] Importantly, the code established a universal ethical framework for clinical research—and its first principle states that "the voluntary consent of the human subject is absolutely essential" to ethical research (Nuremberg Code, 2013). Notably, however, the Nuremberg Code breaks somewhat with the earlier German guidelines, as it emphasizes *voluntariness* over *informedness*—the first principle of voluntary consent prescribes that the subject should be situated to "exercise free power of choice." This emphasis on voluntariness no doubt stems from the direct concern that victims of Nazi experimentation were captive and brutally coerced into medical and other experiments.

---

[2] Many scholars wrongly attribute the origin of informed consent to the Nuremberg Code—as we show above, the germinal idea is present in German experimentation guidelines as early as 1931. It is a tragic irony that the country that enacted some of the first ethical guidelines for human experimentation was forced to confront these issues again on an international scale.

Despite the prioritization of voluntariness, the code does note that consent should entail "sufficient knowledge and comprehension of the elements of the subject matter involved, as to enable [the subject] to make an understanding and enlightened decision" (Nuremberg Code, 2013, n.p.). The code also begins to detail specific kinds of information that should be presented during the consenting process: "the nature, duration, and purpose of the experiment; the method and means by which it is to be conducted; all inconveniences and hazards reasonably to be expected; and the effects upon his health or person, which may possibly come from his participation in the experiment" (Nuremberg Code, 2013). Though voluntariness is foregrounded, the code broke new ground in detailing specific kinds of information that underwrite the idea of "sufficient knowledge and comprehension" in the consenting process.

### *1950s: The Foundations of Informed Consent in U.S. Law and Policy*

*Salgo v Leland Stanford etc. Board of Trustees* (1957) is the first case often cited as establishing the legal doctrine of informed consent for medical practice and biomedical research in the United States (for example, Nelson-Marten & Rich, 1999; Faden & Beauchamp, 1986, p. 57).[3] In the case, the plaintiff was awarded damages for not receiving full disclosure of facts relevant to a particular surgical procedure. The arguments hinged upon whether or not the defendant (doctors) had a duty to disclose all facts including surgical risks and alternative options that might be necessary for a patient to form an "intelligent consent" (Salgo v. Leland, 1957). Notably, the case introduced both the phrases "intelligent consent" and "informed consent" (in that order) to cover the ability of a patient to make a decision based upon full disclosure of facts, risks, and options.

---

[3] Though this paper focuses on the U.S. case, similar developments took place elsewhere. For a history of consent and UK research policy, for example, see: Miller & Boulton, 2007.

But, while *Salgo* is an important case, the roots of informed consent as commonly understood in the United States are more accurately traced to the National Institutes of Health (NIH) some years earlier. In 1953, the NIH opened the Clinical Center research hospital and created one of the first human-subject review boards – the Clinical Center Research Committee (CRC) – to oversee the activities of intramural research programs (Stark, 2011, p. 75).* (*It should be pointed out that a few other universities and hospitals grappled with the same challenges as the NIH Clinical Center around the same time, establishing other, similar committees throughout the 1950s and 1960s. However, the NIH's Clinical Center review committee specifically became the model for later laws and policies governing extramural research and, later, Institutional Review Boards (Stark, 2011).) Further, the hospital aimed to revolutionize the way clinical research was conducted by placing pathology and chemistry labs directly next to hospital wards, allowing for "bench-to-bedside" and "translational research" (Stark, 2011). The hospital also departed from routine clinical research by including experimentation on "normals"—healthy patients either 1) intentionally exposed to infection or treatment or 2) used as a point of comparison for treatments being tried on sick patients.

The internal debate over research ethics prior to the formation of the NIH's Clinical Center is well documented in Laura Stark's book *Behind Closed Doors: IRBs and the Making of Ethical Research*. Stark's work lays bare the conflicted policymakers and tussles between legal counsel and medical staff that informed the development of these early ethical guidelines. Though a more detailed analysis of these conflicts is outside the scope of this paper, it is

important—especially when looking back through policy documents and the development of "informed consent" at the NIH—to take into account the following: 1) the Medical Board was charged with governing many different practitioners (scientists, medical doctors, psychologists, and beyond) whose work was already influenced by established and competing professional codes of ethics and practice, so the Board was hesitant to further burden these individuals and 2) legal counsel to the Clinical Center, Edward Rourke, was concerned that ethics codes were specific to professions and not to places (i.e., NIH Clinical Center) and he continually pushed for legal documentation for over a decade in order to protect the institute from liability (Stark, 2011, p. 105-107).

In February 1953--just prior to the opening of the Clinical Center--the Medical Board released a memo that offered practical guidance on conducting ethical research (Stark, 2011, p. 106). In a section titled "Information for Patient," the memo notes that "each prospective patient will be given an oral explanation in terms suited for his comprehension, supplemented by general written information or other appropriate means, of his role as a patient in the Clinical Center, the nature of the proposed investigation and particularly any potential danger to him" (Medical Board of the Clinical Center, 1953, n.p.). Under the heading "Patient Understanding and Agreement," the memo also notes that "voluntary agreement based on informed understanding shall be obtained from the patient" and stipulates that in all cases "a notation shall be made on the patient's chart of the essential points of the explanation and of the agreement obtained, together with any comment or problems raised by the patient" (Medical Board of the Clinical Center, 1953, n.p.). The policy only required a voluntary signed statement if the procedure involved "unusual hazard." It was not until the final paragraph of the four-page memo that these procedures and responsibility are summed up in a tidy new phrase: informed consent.[4][5]

*1960s & 1970s: Global Developments, Further Codification, and the Belmont Report*

In 1964, the 18th World Medical Assembly adopted the Declaration of Helsinki to establish worldwide guidance for doctors engaged in clinical research. Though amendments in subsequent decades reflect use of the phrase "informed consent," the 1964 version stipulates that clinical research may not take place without a subject's "free consent after he has been informed" (Williams, 2008, p. 652). Like its predecessor the Nuremberg Code, this formulation of consent prioritizes the voluntary (or, in this case, "free") nature of the transaction, but still makes some kind of transfer of information a clear step in the process. The Declaration also stipulates that consent should be obtained in writing as a rule, making it more explicit than many other guidelines, many of which were ambiguous about oral or written consent forms and how to appropriately document the consent process.

In July 1966, the Surgeon General of the United States issued a directive on human experimentation that applied to all grants and awards made by the Public Health Service (PHS), thus expanding federal policies regarding research ethics beyond the NIH to also apply to researchers at non-federal facilities and institutions (U.S. Surgeon General, 1966).[6] The directive cites a PHS resolution from December 1965 stating that support for external research was contingent on ethical review, including "appropriateness of methods to secure informed consent." Further, the directive stated that award grantees were required to keep written records of prior review, in addition to "keep[ing] documentary evidence of informed consent relating to investigations carried out with PHS financial support." It is notable that this directive contains no description of what should be included in (or what constitutes) informed consent, but rather simply dictates that the process be documented, giving rise to the idea of the "consent form."

---

[4] The full paragraph regarding "Responsibility [of the physician]": "He shall be responsible for incorporating in the medical record the information given to the patient and the nature of the *informed consent* or agreement accomplished with the patient, including any comments, objections or general reactions made by the patient" (Medical Board of the Clinical Center, 1953, n.p.).

[5] It is likely the term "informed consent" may be found in internal NIH memos (as indicated by Stark, 2011), but the important takeaway is that the phrase did not originate within the context of clinical research with *Salgo* as is usually attributed (this memo predates the case by at least four years).

Further clarity on ethical review of external research supported by federal grants and awards was provided by the U.S. Department of Health, Education, and Welfare (DHEW) in 1971.[7] In "The Institutional Guide to DHEW Policy on the Protection of Human Subjects," informed consent is defined as "the agreement obtained from a subject, or from his authorized representative, to the subject's participation in an activity" (U.S. Department of Health, Education, and Welfare, 1971). The document detailed six precise elements of informed consent:

1) A fair explanation of the procedures to be followed, including an identification of those which are experimental;

2) A description of the attendant discomforts and risks;

3) A description of the benefits to be expected;

4) A disclosure of appropriate alternative procedures that would be advantageous for the subject;

5) An offer to answer any inquires concerning the procedures;

---

[6] There were actually two memos (February and April 1966) which superseded this memo. It appears that the memos were issued with increasing policy strength—meaning they applied first only internally and then externally to all grants and awards of the public health service.
[7] The Department of Health, Education, and Welfare (DHEW) was renamed in 1979 to Department of Health and Human Services (HHS).

> 6) An instruction that the subject is free to withdraw his consent and to discontinue participation in the project or activity at any time (U.S. Department of Health, Education, and Welfare, 1971)

This marked the first time in United States policy or case law that the contents of the information to be given in informed consent are explicitly defined.

Three years later the United States Congress passed the National Research Act (PL 93-348) that would eventually be codified into 45 CFR 46, better known as "The Common Rule." The original version of the National Research Act actually contains very little specific guidance for the treatment of human subjects in research. Importantly, however, the law established the National Commission for the Protection of Human Subjects of Biomedical and Behavioral Research—the commission eventually responsible for authoring *The Belmont Report on Ethical Principles and Guidelines for the Protection of Human Subjects of Research*. The law tasked the commission with conducting a comprehensive study of ethical principles regarding human subjects research, including "the nature and definition of informed consent in various research settings" and later to "identify the requirements for informed consent to participation in… research by children, prisoners, and the institutionalized mentally infirm" (U.S. Congress, 1974). It also required funded research to be reviewed by formal Institutional Review Boards, though it makes no explicit mention of informed consent procedures or requirements.

The national commission met in 1976 and the Belmont Report was published in 1979. Today, the report and its three major ethical principles—respect for persons, beneficence, and justice—are regarded as foundational for ethics in human subjects research. Under respect for persons, the report states that "individuals should be treated as autonomous agents" and that subjects should enter research "voluntarily and with adequate information," harking back to some of the earliest formulations of consent in medical settings (National Committee, 1979, n.p.). In line with later, more explicit codifications of consent processes, however, the report offers specific guidance on how to apply the principle of respect for persons through the process of informed consent. Focusing on the informational element more than consent itself, "adequate information" is said to include research procedures, purposes, risks and anticipated benefits, alternative procedures in cases of biomedical research, and statements welcoming questions or withdrawal from the research at any time. Further, the report addresses aspects of the presentation of information which may adversely affect a subject's ability to make an informed choice (for example, that information should not be presented in a disorganized or rapid fashion). Finally, the report stipulates that it may be necessary in some cases to conduct written tests or oral examinations to ensure subjects are informed. Though these informational and administrative dimensions are discussed in detail, it is worth noting that the report does acknowledge the importance of voluntariness, making it clear that consent is only valid if it is not coerced or unduly influenced.

**Current Landscape: Past, Present, and Emerging Challenges**

*Current Policies Regulating Informed Consent*

Today, the use of human subjects is still regulated in the United States by the Common Rule.[8] Many academic institutions in the United States also sign federalwide assurance agreements that make all human subjects research subject to review by an Institutional Review Board whether or not it is funded in any capacity by the federal government. As a result of these agreements, academics that partner with researchers in private industry are often still obligated to obtain ethical review at their home university.[10]

---

[8] It is called the "Common Rule" because it applies to "all research involving human subjects conducted, supported or otherwise subject to regulation by any federal department or agency." Further, "research… otherwise subject to regulation" encompasses activities that fall under specific responsibilities of a regulatory agency like the Food and Drug Administration's (FDA) requirements for investigational new drugs or medical devices. Other agencies and jurisdictions, like the the Federal Trade Commission's (FTC) regulation of unfair and deceptive business practices, may incidentally have oversight over some private sector research as it relates to consumer business practices but are separate from the Common Rule.

Research is defined by the Common Rule as a "systematic investigation, including research development, testing and evaluation, designed to develop or contribute to generalizable knowledge" (45 CFR 46 §4.6102 - Definitions). When research falls under the regulatory purview of the Common Rule, it must meet specific stipulations with regard to informed consent. Specifically, section 46.116 (as amended in 2009) establishes the general requirements for informed consent currently codified into United States law, requiring that "no investigator…involve a human being as a subject in research covered by this policy unless the investigator has obtained legally effective informed consent of the subject or the subject's legally authorized representative." Moreover, the law specifically states that "information given to the subject must be in language understandable to the subject or representative, and no exculpatory language may be included that waives the subject's legal rights or the investigator's liability for negligence."

---

[9] As of the writing of this article, there are efforts underway to revise the Common Rule to better accommodate non-medical research and, in particular, data scientific research. However, the proposed changes (which would streamline review for certain projects and potentially exempt wide swaths of "big data" research that may still implicate human subjects) have been controversial—the Department of Health and Human Services noted that they received over 2100 public comments on the proposal (Malakoff, 2016). In light of the tentative and uncertain future of these proposals—and to not distract from our historical analysis—we have omitted any extensive discussion of the proposed changes. (For a discussion of the proposal and its potential shortcomings, see: Metcalf & Crawford, 2016.)

[10] It should be noted, however, that there is a known loophole where protocol may be segmented so that academic researchers not dealing with data collection may be able avoid IRB review at their home institutions and industry partners are thus exempt and able to freely conduct data collection on human subjects. This loophole was used in the Facebook emotional contagion study (Carberry, 2014).

According to U.S. statutory authority, basic elements of informed consent when seeking informed consent from subjects today should include the following: 1) a statement that the study involves research, the purposes of research, expected duration of subject participation, descriptive procedures to be followed, and identification of experimental procedures; 2) description of any reasonably foreseeable risks or discomforts to the subject; 3) description of benefits to the subject or others reasonably expected from this research; 4) disclosure of any alternative procedures or treatments that could be advantageous to the subject; 5) statement regarding the extent (if any) confidentiality of records identifying the subject will be maintained; 6) if risk is more than minimal, an explanation of other medical treatments available if injury occurs; 7) contact information for questions about the research or research subject's rights, or whom to contact in the event of research-related injury; and 8) a statement that participation is voluntary and refusal or discontinuation of participation will not result in penalty or loss of benefits the subject is otherwise entitled (45 CFR 46 § 46.116 (a)). § 46.117 also mandates that informed consent be documented through a written consent form approved by an Institutional Review Board, and that the form may be read and signed by the subject or presented orally as long as a witness is present (unless a waiver has been granted). There is also a somewhat implied link between Institutional Review Board review and the requirement of informed consent practices.

***Past and Ongoing Challenges to Informed Consent***

The highly formalized and bureaucratic nature of these requirements—along with their clear indebtedness to medical research contexts—has made informed consent controversial in broader social scientific and other research contexts. Translating ethical requirements from physical interventions (like, for example, using a needle to inject a drug into a person's body) to other kinds of research methods (like relatively unintrusive observation of individuals outside of a laboratory) has proven a challenge. Data analyses, in particular, "rarely appear as a direct 'intervention' in the life or body of an individual human being" (Metcalf & Crawford, 2016, p. 3). It is also unclear how to assess harm or potential risk outside of physical interventions or in-person interactions, as with the possibility of reidentified data or harms that may occur on extended timescales (e.g., unlike physical interventions, which take place in discrete times and places, data may be analyzed years after the point of collection and reveal new knowledge and generate new risks for an individual). Given the mismatch in contexts, scholars and researchers in the humanities, social sciences, and other domains have been reluctant to embrace the often legalistic and administrative version of ethical review set out by IRBs (Schrag, 2010), while researchers in computer science or mathematics have escaped IRB regulation almost entirely (Metcalf & Crawford, 2016, p. 3-4).

Nonetheless, ethical review boards like IRBs have—since the 1970s, and particularly since the 1990s—been increasingly aggressive when it comes to regulating non-medical, non-experimental research (Schrag, 2009). But for many qualitative researchers the "standard and formalised practices [of informed consent] do not sufficiently accommodate the social context in which qualitative research is now carried out" (Miller & Boulton, 2007, p. 2204). For example, the practice of fully disclosing the risks of research and obtaining explicit, written consent at the very start of a research project does little to promote ethical conduct on the part of researchers embedded in evolving field sites during long-term projects, where consent can range from explicit to implied and must be negotiated constantly over time. In these cases, a researcher's ongoing sense of responsibility toward her research subjects is more important than one-time consent at the start of a study (Miller & Boulton, 2007, p. 2204). As Newmahr & Hannem (2016) flatly note, "ethnographers cannot obtain informed consent beforehand from everyone with whom we will speak. We don't yet know who they are, what we want to talk about, or in what settings those conversations might happen" (p. 4). Moreover, simply making adjustments to please an ethical review board and "check off" administrative requirements in full view of their impracticality puts researchers in the potentially duplicitous position of acting one way for their IRB and another way in the field.

During the 1990s and 2000s, the subdomain of Internet Research Ethics began to confront the problems of consent in digital environments, where "lurking" (i.e., the possibility of observing participant behavior without making one's presence known) is possible and public/private distinctions are not easily drawn (see, for example: Kraut et al., 2004; Buchanan & Ess, 2009; Markham & Buchanan, 2012). Even when participants may be identified and contacted, a lack of researcher and subject interaction renders the effective assessment of the participants' comprehension of the form or potential risks more difficult. Though some recommendations suggest adding affirmative "click to accept" buttons on digital consent forms to help promote informedness, this only serves to further highlight the limitations of consent forms within new, online research contexts (Kraut et al., 2004). Further, there are additional logistical challenges associated with obtaining consent from research conducted online, especially when the subject pool is extremely large and/or when a dataset is publicly accessible and individuals contained in the data are difficult to track down (Vitak, Shilton, & Ashktorab, 2016).

Beyond the practical, the legalistic and administrative nature of informed consent also occupies a tenuous conceptual place in non-medical research, as it "presumes a great deal—not only about the nature of ethical research but also about the nature of research itself" (Bell, 2014,

p. 517). For one, implicit in this model is the idea of research participants "as rational, individual, modernist—and Western—subjects for whom the concept of informed consent and its documentation by a signed form is unproblematic" (Miller & Boulton, 2007, p. 2204). Discussing issues of informed consent in anthropological research, Bell (2014)—drawing on Jacob and Riles (2007) and others—argues that the "totalizing views" of researchers that are enacted by research ethics documents and practices ultimately reify the very inequalities between supposedly driven and ethically-blind researchers and passive, vulnerable research subjects they purport to address (p. 518).

### *New Challenges and Changing Research Contexts*

The tension between non-medical researchers and the demands of informed consent is made more complicated today by the proliferation of sensors, "smart" objects, social networking sites, and massive social data repositories—all of which are inherently about the people who operate or are contained within these systems. The study of these information systems have forced regulators and ethicists alike to reconsider the human subject in computational, statistical, and data scientific research. There are growing opportunities in "big data" research to uncover new knowledge or make more personal (and perhaps more accurate) inferences, not to mention new opportunities to interact with and experiment on individuals through feedback or manipulation of online environments and platforms. And academics are not the only ones poised to benefit from these developments--research on data about human subjects has moved outside of the academy, as private firms and technology companies increasingly invest in internal research to develop products and engage users and develop deeper knowledge about human behavior. For researchers in academia and industry alike, understanding fundamental aspects of habit formation, resource consumption, social engagement, and other activities is central to successfully integrating technology into daily use and for broader societal benefit.

As industry attention to fundamental or generalizable knowledge about human behavior and sociality grows, the distinction between conventional or academic social scientific research and industry- or market-oriented research has become less clearly delineated (boyd, 2016). At social media company Facebook, for example, academic research is an important part of the company's culture, which includes an active data science team with extensive ties to academia (Grimmelmann, 2015, p. 222; see also, Facebook, 2016). Other companies—from microblogging site Twitter to dating site OKCupid to transportation network Uber—readily advertise and share details regarding their research (Twitter, 2016; OKCupid, 2016; Uber Design, 2015).

Additionally, part of the promise and underlying mythology of big data and a world instrumented with the Internet of Things (IoT) is to uncover new insights, cover multiple use cases, and may be linked to many other products and services within the same company. Alternatively, future acquisitions and consequential merging of user data makes it difficult to assess how seemingly independent information streams could merge in the future. The increasing number of user interfaces, information, and opportunities to interact with different aspects of individuals' lives comes with unprecedented potential to study our public and private lives. These challenges have been particularly evident in controversial online experiments that sparked widespread conversations and concerns—in particular, Facebook's emotional contagion experiment (Kramer, Guillory, & Hancock, 2014) and OKCupid's match ranking experiment (Rudder, 2014). The Facebook study, in particular, stirred a great deal of debate—both in the public and in academia—around the ethics of an experiment designed to emotionally manipulate users without consent or a post-experiment debrief (Puschmann & Bozag, 2014; Gray, 2014; Flick, 2016).

In the United States, research in private industry that is not supported by federal funds—like the study carried out by Facebook—generally falls outside of federal research regulations that would require ethical review or consenting processes. This is largely a result of the historically economic and self-interested (i.e., non-generalizable) motivations of research conducted in industry, where A/B testing, usability studies, and other methods are commonly used as methods to improve products and services for the benefit of users and profit margins. Current investigations into how users engage with products and use technology—which are often systematic and conducted by highly educated employees or partner organizations—blur this historically distinct line. As big data methods drift toward not simply uncovering user patterns with a particular product but toward generalizable insights about human behavior--either through deeply intimate inferences derived through observational study or experimental manipulation—there is a distinct shift in the intent and opportunity for internally conducted research.

But private industry's move toward questions surrounding generalizable knowledge about human behavior raises new questions as to how the ethics of research on human subjects should apply in these environments. Of particular interest are new challenges to the protection of individuals and groups, as large scale industry-driven behavioral research and development threatens to undermine privacy and compromise the security of users' data (Lohr, 2013; Schneier, 2016). Currently, public and private organizations alike are left without robust guidance on how to responsibly execute research and development using potentially sensitive datasets.[11] As a result, organizations have been left with research ethics frameworks and legacy processes that are poorly suited for modern data analyses, often relying instead on imperfect notice-and-consent policies to guide the collection and use of data, regardless of whether or not that data will be used for research. New discussions within industry have shown some attempts at self-regulation through internal review processes, like Facebook's recently developed internal review, which places research oversight into existing organizational infrastructure and offers multiple opportunities to flag particular projects for review (Jackman & Kanerva 2016). Whether or not processes like Facebook's ultimately promote ethics is an ongoing question (Hoffmann, 2016), but they nonetheless highlight the challenges to adapting previous principles and guidelines for an unregulated and untested private sector context (Polonetsky, Tene, & Jerome, 2015).

Developments like the Facebook review process are, however, only beginning to emerge. Up until now, however, the lack of clear or explicit research ethics guidance for data science and industry research created a policy vacuum that has largely been filled by legal mechanisms of privacy policies (or similarly, notice and consent procedures for data collection and use), legal mechanisms that developed independently from (and have little in common with) informed consent in research policy. Unfortunately, commentators and scholars alike often conflate "informed consent" with notice and consent procedures, and many legal discussions of informed consent online focus on improving the informativeness and readability of privacy policies in an effort to make them function more like informed consent in research (see, for example: Pollach 2005; Friedman, Felten, & Millett, 2000). Other scholars have concluded that the kind of consent represented by click-through agreements like privacy policies are insufficient for promoting and protecting users (Custers, Schof, & Hermer, 2014; Barocas & Nissenbaum, 2014).[12]

---

[11] Many companies do not label internal practices formally as "research and development," but these activities may lead to internal product or platform development as well as generalizable knowledge (e.g., about human behavior) in some cases.

**Lessons from the History and Current Landscape of Informed Consent**

Not only is research ethics about orienting ourselves towards research and data subjects in particular ways, but it is also about interpreting and applying ethical ideals in a non-ideal world. In the preceding sections, we have shown how informed consent—in its legal and political history—has evolved alongside the social, political, and research-related contexts it is intended to regulate, from the earliest formulations of voluntary consent to the explicit codification of informed consent as an expression of respect for persons in The Belmont Report. Moreover, the detailed policy history showed how different social and political factors—be they dramatic, as in the case of the Nuremberg Code, or seemingly boring, as in the administrative minutiae of the early NIH—help shape what "consent" means in practice. In some way, then, this account makes explicit the historical processes that underwrite, as Jacob Metcalf (2015) calls it, the "hard-won" social trust that affords physicians and clinicians the distinction between practice and research

---

[12] The limits of privacy policies or mere notice and consent procedures stem, in part, from differences in origins, theoretical grounding, and structural elements of informed consent. The core ethical principles presented by the Belmont Commission—in particular, respect for persons—drew on robust philosophical debate within the Commission and were operationalized in law by drawing upon legal traditions within tort law (Faden, 1986). As a result, these regulations appropriately considered the context of research settings and actively interwove moral theory and extant legal traditions. By contrast, online privacy policies evolved pragmatically in the mid-to-late 1990s as way for private companies to stave off further regulation (Solove & Hartzog, page 593). Consequently, the elements contained within privacy policies or the outgrowth of notice and consent procedures pertaining to any data collection online and offline are driven by case law and regulatory precedent, and largely focus on protecting companies' private interests by legitimizing almost any collection, use, or disclosure once consent has been obtained, instead of empowering individuals to self-manage their privacy (Solove, 2013). Moreover, research ethics regulations must be understandable to broad audiences of stakeholders—from researchers, graduate students, ethicists, lawyers, and even the public. Privacy policies, on the other hand, have been described as "written by lawyers for lawyers" (Cate & Mayer-Shönberger, 2013).

(n.p.). By consistently responding to ethical breaches and evolving alongside the scientific, technological, and administrative contexts that shape research and practice, conventional medical and behavioral research ethics has demonstrated a certain kind of commitment to important human values like respect and beneficence. Comparatively, today's data scientists and software engineers do not have the same kind of history of struggle and organized response to fall back on (Metcalf, 2015). Based on the history and challenges presented here, we summarize three key lessons that should inform future work and discussion on respect and consent online and for data science today.

1. Though the idea and the legalistic practices surrounding informed consent are often conflated, it's important to note that the idea and value of informed consent precedes our current understanding of it as an administrative or bureaucratic process. At various points in its early history, informed consent functioned as an often informal way of promoting or respecting the autonomy of individual subjects. Often, early manifestations of the practice emphasized: A) voluntariness on the part of subjects and B) mutual agreements reached informally between subjects and medical practitioners or early researchers. Though one might rightly question the power dynamics present in some of these early scenarios, we nonetheless can argue that there are no strict policy reasons for consent to take the bureaucratic, administrative process of recordkeeping with which it has become synonymous. There is no value limitation keeping the consenting process as a signed agreement—future consent could be obtained through audio interaction or other interactive modalities.

    Moreover, we have provided evidence—through a close reading of the relevant United States policy history—of how the bureaucratization of informed consent took root

and eventually solidified. Formal documentation and routinized information practices did not become common until the latter half of informed consent's more than 100 year policy history and their rise was intimately bound up with the institutional and bureaucratic structures of the NIH in the 1950s. During that time, a few influential players—most notably Edward Rourke, the Clinical Center's legal counsel—were instrumental in laying the foundation for the idea that research ethics had more to do with administrative requirements and institutional liability than actually promoting respect for persons. In the mid-1960s, the Clinical Center's model for ethical oversight was expanded beyond the NIH to apply to all grants and awards made by the Public Health Service (PHS) to researchers at non-federal facilities and institutions—this expansion was subsequently codified in 45 CFR 46 (the Common Rule). In summary: the shape and scope of informed consent today was not an accident, but was instead shaped by particular political decisions made starting in the 1950s. Written documentation with ever evolving boilerplate language is not the only way to obtain informed consent, but it was a scalable policy for national research designed to fit some of the ethical dilemmas from that time.

2. Expanding on Metcalf's (2015) discussion of the "hard-won" social trust that the history of research ethics debates in biomedical contexts has afforded the public and researchers alike, we must also recognize the robust history of pushback and ethical debate among social scientists and humanities researchers in the academy. Though the tension between institutional ethical review and these types of research is hardly resolved, there is a decades-long history of discussion and debate that has helped to inform and further develop the ethical identities and commitments of specific academic disciplines—a history that computational, "big data" social scientific research in technology industry settings is not able to also fall back on. But where Metcalf (2015) and Metcalf and Crawford (2016) understandably place the origins of these struggles over research ethics with Nazi experimentation and the Nuremberg Code, we show how informed consent, in particular, has an even longer history, with its earliest manifestations forged in debates over medical treatments and individual autonomy nearly 50 years prior. Ironically, these early controversies culminated in robust ethical guidelines for researchers in Germany in 1932—guidelines that were ignored by the very Nazi officials and doctors whose prosecution and conviction would lead to the development of the Nuremberg Code. Though this may appear on the surface at historical nitpicking, it serves to deepen and extend the argument that researchers in math, computer science, and data science do not have a similarly robust history of research controversy and debate to which they can appeal. Consequently, calls to throw out considerations of informed consent (or, for that matter, other pieces of research ethics) that do not heed this history come across as, at best, hollow and, at worst, self-interested.

3. The political history of informed consent reveals the importance of context for shaping how we conceive of and implement protections for certain key ethical values, like respect for persons. For example, the early focus on the *voluntariness* of consent reflected the prevailing concern with with placing limits on coercive state power. The emphasis on voluntariness is made particularly sharp in the Nuremberg Code and its response to the horrors of Nazi experimentation—here, realizing respect through consent hinges on subjects being free from powerful institutional and political forces that dominate, coerce, and instill fear. In view of these conditions, voluntariness (and, in particular, the absence of coercion) is central to appreciating the dignity of persons. However, as research came to be formalized within bureaucratic and administrative structures like modern universities and the NIH, however, the idea of realizing respect through consent shifted accordingly. Moving from an emphasis on voluntariness to *informedness*, informed consent in administrative and legalistic context hinges on the presence or absence of certain kinds of information and certain levels of understanding. The rise of "informedness" as central to consent can be traced to its development within the NIH and their use of informational and administrative practices necessary for accountability and liability purposes. The mandate to present particular information (e.g., risks and benefits, alternative options, explicit ability to opt out, etc.) also played into earlier emphasis on voluntariness in that individuals would not be empowered to make a truly autonomous decision without such material information available.

Today, we must pay close attention to the features of specific research contexts and rethink the types of coercive power with which we might be concerned. While abuses in the academy are still very much a concern, further discussions of respect for persons in data-intensive, digital industry contexts must take seriously what it might mean to protect individuals and groups from the increasingly pervasive influence of online platforms and technology companies. Rather than an informed consent that hinges on physical interventions staged or supported by the state, we need a mechanism that takes seriously questions of justice and exploitation under high technology, capitalist firms. We need to ask difficult questions about what it means, to recall language from the Nuremberg Code, to promote the "exercise [of] free power of choice" in online ecosystems increasingly dominated by a handful of large Internet companies interested in controlling the flow of information between users, companies, and third parties like advertisers. Even if solutions to these emerging challenges end up looking different than conventional informed consent, the impracticability of consent when dealing with thousands to even millions of individuals in the form of data points should not lead us to dismiss those concerns, nor should it distract us from heeding the key values (like autonomy or respect for persons) that consent was designed to support in the first place.

As Mary Gray (2014) puts it, our "'ethical dilemmas' are often signs that our methodological techniques are stretched too thin and failing us" (n.p.). In recognition of the limits of informed consent for realizing respect, we advocate for future work that builds on this history by pairing it with careful explication of the value of respect itself, especially as presented by deontological and contractualist philosophers. Not only must careful attention to social, legal, and political history inform research ethics and practice, but the assumptions that underwrite our ethical commitments must continually be revisited in light of new and emergent social and technological possibilities. In addition, we need to think carefully about what it means to realize respect in specific social, political, or technical contexts. This context and political complexity matters for how we think about operationalizing respect as a value for research ethics. It is vital that we pay attention to how certain features of research settings or ICTs can "serve various interests and exclude or marginalize other interests in ways that materially affect the opportunities and limitations of real flesh and blood persons" (Dillon, 2010, p. 25).

**Conclusion**

In the preceding, we have shown how informed consent has not been a static or monolithic mechanism but, rather, has constantly evolved from early human subjects research practices through (and beyond) codified documents like the Belmont Report that set forth ethical standards in national research policy, with particular focus on the United States context. By marrying close legal and policy analysis how informed consent has, at various point in its development, sought to operationalize the ideal of respect for persons. In so doing, we have sought to recover valuable insights relevant to discussions around data, technology, and research today.

In particular, we focused on the situated nature of informed consent policies and the ways in which they evolved alongside the social and political context within which they operated. This situatedness is particularly important for thinking about how to best realize respect in online and industry research contexts. In so doing, however, we do not mean to overemphasize the importance of informed consent—it is, after all, not the only mechanism that might be employed to operationalize respect. We recognize that while a focus on consent may make ethics conversations more manageable, too much focus on it risks losing the full force of respect as a guiding value or research ethics. Or, as Labacqz (2005) summarizes: "As respect for persons is reduced to autonomy and autonomy to self-determination or freedom of choice, the logical outcome is that the broad-ranging principle of "respect for persons" is then truncated into the rule of "informed consent." (p. 101). Rather than frustrate us, then, the history presented in the foregoing should be a lesson that informed consent's legalistic manifestation is not the only way to realize important values like autonomy or respect.

Rather than dismissing informed consent as intractable or not applicable to industry research, we argue that its history has relevance for thinking about how to best realize respect in the context of online experimentation and research of the kind conducted by Internet companies like Facebook and OKCupid. Whether or not respect online takes the form of something more or less like "informed consent" in the future is, we think, a less fruitful approach than recognizing that the history of informed consent in research ethics shows us that respect in online research will require careful attention to the particular affordances, political, and other contextual factors of particular kinds of online platforms. Marrying close historical analysis with careful philosophical explication—and critically interrogating both—is, we argue, an integral step towards the development of a 21st century research ethic and actionable policy recommendations.

**Acknowledgements**

The authors would like to thank two anonymous reviewers, Luke Stark and other commenters from the 2016 Privacy Law Scholars Conference, and participants in the Alan Turing Institute Workshop on The Ethics of Data Science (University of Oxford, 2015) for their helpful feedback at various points in the development of this article. Additional thanks are also due to Paul Duguid and Geoff Nunberg for invaluable feedback and guidance on early research and writing that lead to this project and collaboration. Finally, this project was generously supported by the UC Berkeley Center for Technology Society and Policy (CTSP), UC Berkeley Center for Long-Term Cybersecurity (CLTC), and a National Science Foundation Graduate Research Fellowship. Author names are listed in reverse alphabetical order; both authors contributed equally to this article.